# Toroidal optical transitions in hydrogen-like atoms

Ilya Kuprov[1], David Wilkowski[2,3,4,*], Nikolay Zheludev[2,5,6]

[1]School of Chemistry, University of Southampton; United Kingdom.

[2]Centre for Disruptive Photonic Technologies, The Photonics Institute, Nanyang Technological University; Singapore 637371, Singapore.

[3]Centre for Quantum Technologies, National University of Singapore; 117543 Singapore, Singapore.

[4]MajuLab, International Joint Research Unit UMI 3654, CNRS, Université Cote d'Azur, Sorbonne Université, National University of Singapore, Nanyang Technological University; Singapore.

[5]Optoelectronics Research Centre, University of Southampton; United Kingdom.

[6]Hagler Institute for Advanced Studies, Texas A&M University; USA

*Corresponding author: david.wilkowski@ntu.edu.sg

**Abstract**

It is commonly believed that electromagnetic spectra of atoms and molecules can be fully described by interactions of electric and magnetic multipoles. However, it has recently become clear that interactions between light and matter also involve toroidal multipoles – toroidal absorption lines have been observed in electromagnetic metamaterials. Here we show that a new type of spectroscopy of the hitherto largely neglected toroidal dipolar interaction becomes feasible if, apart from the classical $\mathbf{r}\times\mathbf{r}\times\mathbf{p}$ toroidal dipole density term responsible for the toroidal transitions in metamaterials, the spin-dependent $\mathbf{r}\times\boldsymbol{\sigma}$ term (that only occurs in relativistic quantum mechanics) is taken into account. We show that toroidal transitions are odd under parity and time reversal symmetries; they can therefore be observed and distinguished from electric multipole and magnetic dipole transitions.

**Main text**

The description of electromagnetic properties of matter traditionally involves multipole expansions (*1,2*) that are widely used in the study of biological, chemical, atomic, and subatomic phenomena where electromagnetic processes are described by electric and magnetic multipoles arising from moving charges and current loops. Toroidal multipoles (*3*), generated by currents flowing on toroidal surfaces, constitute the rarely acknowledged third independent family of vector potential sources; they provide distinct and significant contributions to electromagnetic properties of matter.

Static toroidal dipoles (Fig. 1, top right) were introduced in 1957 by Zeldovich (*4*); they are found in magnetic materials (*5*), atomic nuclei (*6*), single molecule magnets (*7*), fullerenes (*8*) and in solid state physics (*8*). Dynamic toroidal dipoles with oscillating currents interact with oscillating electromagnetic fields, and thus contribute to optical properties of matter. They were first observed in artificial metamaterials (*9*); in recent years, they made appearances in various forms of artificially structured matter across the electromagnetic spectrum (*10*): nanoparticles (*11*), discussions of noise-resistant quantum devices (*11*), anapoles (*6,12*), dark matter (*13-15*), communication using Aharonov-Bohm effect (*16,17*), electromagnetic reciprocity (*18*) and as contributors to non-radiating charge-current configurations (*16,19*).

Because the work on dynamic toroidal multipoles has so far mainly involved Maxwell electromagnetic metamaterials, the associated spin effects have been overlooked (*9,19*). However, those are unavoidable in quantum mechanical systems, such as atoms and molecules – Dirac's equation only conserves the sum of angular momentum and spin (*20*). In this communication, we consider the possibility of using spin effects to open new toroidal excitation channels in quantum emitters (QE). We demonstrate that photonic excitation of QE based on the non-relativistic toroidal dipole operator $\mathbf{r}\times\mathbf{r}\times\mathbf{p}$ is exceedingly difficult due to field frequency-curvature mismatch and inconspicuous selection rules. However, the appearance of the spin part $\mathbf{r}\times\boldsymbol{\sigma}$ in the relativistic case eliminates the frequency-curvature problem and yields distinctive selection rules. We identify pairs of energy levels and experimental conditions for which toroidal transitions in hydrogen-like atoms are likely to be detectable.

**RESULTS**

**Non-relativistic case: vector potential curvature problem**

The length scale of electronic optical transitions of interest in atomic physics, chemistry, and molecular biology is typically two to four orders of shorter than the wavelength of light. This creates a problem for a hypothetical spectroscopy based on the classical $\mathbf{r}\times\mathbf{r}\times\mathbf{p}$ toroidal dipole – we demonstrate here that significant electric field curvature is required on the molecular length scale, and that the optimal frequency-wavelength combination for $\mathbf{r}\times\mathbf{r}\times\mathbf{p}$ spectroscopy is three orders of magnitude away from the constraint imposed by the speed of light.

For an electron in a scalar potential $\varphi(\mathbf{r})$ and a vector potential $\mathbf{A}(\mathbf{r})$, the Hamiltonian, in the Coulomb gauge, can be split into the background term $\hat{H}_0$ and the coupling term $\hat{H}_1$ (*1*):

$$\hat{H}_0 = \frac{\mathbf{p}^2}{2m} - e\varphi + \frac{e^2}{2m}\mathbf{A}\cdot\mathbf{A}, \qquad \hat{H}_1 = \frac{e}{m}\mathbf{A}\cdot\mathbf{p} \qquad (1)$$

where $m$ is electron mass, and $e$ is the elementary charge. The two electromagnetic operators appearing in $\hat{H}_0$ are not interesting to us here – they are purely coordinate operators that cannot contain anything proportional to $\mathbf{r}\times\mathbf{r}\times\mathbf{p}$.

In the coupling term $\hat{H}_1$, consider an electromagnetic plane wave with a vector potential amplitude $\mathbf{A}_0$ and angular frequency $\omega$, travelling in the direction of a unit vector $\mathbf{n}$:

$$\mathbf{A}(\mathbf{r},t) = \mathbf{A}_0 \exp\left[i(\omega/c)\mathbf{n}\cdot\mathbf{r} - i\omega t\right] \qquad (2)$$

The toroidal operator originates in the Taylor expansion of the spatial part:

$$\exp\left[i(\omega/c)\mathbf{n}\cdot\mathbf{r}\right] = 1 + i(\omega/c)\mathbf{n}\cdot\mathbf{r} - (\omega/c)^2(\mathbf{n}\cdot\mathbf{r})^2/2 + \ldots \qquad (3)$$

for which straightforward vector calculus transformations yield:

$$(\mathbf{n}\cdot\mathbf{r})^2\mathbf{p} = \mathbf{r}(\mathbf{r}\cdot\mathbf{p}) - \left((\mathbf{n}\times\mathbf{r})\cdot(\mathbf{n}\times\mathbf{r})\right)\mathbf{p} - \mathbf{r}\times\mathbf{r}\times\mathbf{p} \qquad (4)$$

Although electronic transitions under $\mathbf{r}\times\mathbf{r}\times\mathbf{p}$ operator are possible – a direct calculation confirms that there are plenty of non-zero transition moments within the orbital structure of any reasonable molecule – the conclusions regarding the sensitivity and selectivity of such transitions are pessimistic. Firstly, there are electric and / or magnetic operators in every irreducible representation of the rotation group, meaning that any allowed $\mathbf{r}\times\mathbf{r}\times\mathbf{p}$ transition will overlap with an electric or a magnetic transition of exactly the same frequency. Secondly, $\mathbf{r}\times\mathbf{r}\times\mathbf{p}$ term in the plane wave expansion in Equation (3) is a factor of $(\omega r/c)^2 = (2\pi r/\lambda)^2$ – here $r$ is the characteristic size of the molecular orbitals in question – weaker than the zero-order term responsible for the electric dipolar interaction, meaning that the transition rate in the Fermi golden rule is a factor of $(2\pi r/\lambda)^4$ smaller. For common atomic diameters (H: 0.25 Å, Cs: 0.52 Å) and a wavelength in visible or near infrared, we expect a typical $\mathbf{r}\times\mathbf{r}\times\mathbf{p}$ transition to be a factor of $10^5$ to $10^6$ weaker than a typical electric dipolar transition.

These are likely the reasons why toroidal transitions have never been observed in chemical spectroscopies. So far, dynamic toroidal excitations have only been seen in metamaterials, where resonant structures with the feature size comparable to the wavelength can be engineered in such a way that $2\pi r/\lambda \approx 1$, and the lower order electric and magnetic multipoles suppressed by design (*9,19*).

**Relativistic case: spin part of the toroidal operator**
We now consider an important difference between Maxwell electromagnetic metamaterials and molecules: the angular momentum $\mathbf{r}\times\mathbf{p}$ that occurs in the $\mathbf{r}\times\mathbf{r}\times\mathbf{p}$ is only a constant of motion in the rotation subgroup of the Galilean group. The corresponding invariant in Lorentz and Poincare groups also includes relativistic boosts, leading to the conservation of the sum of angular momentum and spin. The corresponding $\hbar(\mathbf{r}\times\boldsymbol{\sigma})$ correction to the toroidal operator, where $\boldsymbol{\sigma}$ is a vector of Pauli matrices, deserves a good look.

In the standard derivation of the electromagnetic and spin Hamiltonian by approximate elimination of negative energies from Dirac's equation (*2*), the first occurrence of the $\mathbf{r}\times\boldsymbol{\sigma}$ product is in the angular magnetoelectric (AME) part (*21*) of the gauge invariant form of spin-orbit (SO) coupling:

$$\hat{H}_{\text{SO+AME}} = -\frac{e\hbar}{4m^2c^2}\boldsymbol{\sigma}\cdot\left[[\nabla\varphi]\times(\mathbf{p}+e\mathbf{A})\right] \quad (5)$$

The conventional spin-orbit term is not interesting to us here – it comes from the scalar potential – but the triple product with the vector potential is easily rearranged into a form that exposes the presence of the spin part of the toroidal operator:

$$\hat{H}_{\text{AME}} = \frac{e^2\hbar}{4m^2c^2}\mathbf{A}\cdot\left[[\nabla\varphi]\times\boldsymbol{\sigma}\right] \quad (6)$$

This is directly visible in the special case of a hydrogen-like atom with a nuclear charge $Z$:

$$\nabla\varphi(r) = -\frac{Ze}{4\pi\varepsilon_0}\frac{\mathbf{r}}{r^3} \quad (7)$$

where the spin part $\mathbf{r}\times\boldsymbol{\sigma}$ of the toroidal dipole moment is coupled to the vector potential:

$$\hat{H}_{\text{AME}} = -\frac{\alpha\mu_B^2}{ec}\frac{Z}{r^3}\mathbf{A}\cdot[\mathbf{r}\times\boldsymbol{\sigma}] \quad (8)$$

where $\alpha$ is the fine structure constant, and $\mu_B$ is the Bohr magneton. This was previously considered a spin interaction (*21*); the form presented here views it instead as pertaining to the relativistic component of the toroidal dipole moment.

**Selection rules**

Although the interaction described by Eq (8) is exceedingly weak, its transitions are nonetheless expected to be observable because their selection rules are different from the selection rules associated with electric and magnetic dipoles (Table 1). The corresponding operators are:

$$\mathbf{d}\propto\mathbf{r}, \quad \boldsymbol{\mu}_S\propto\mathbf{S}, \quad \mathbf{t}_S\propto\mathbf{r}\times\mathbf{S} \quad (9)$$

where $\mathbf{d}$ stands for electric dipole, $\boldsymbol{\mu}_S$ stands for the spin part of magnetic dipole, $\mathbf{t}_S$ stands for the spin part of the toroidal dipole, and the electron spin operator is $\mathbf{S}=\hbar\boldsymbol{\sigma}/2$. In each specific spin-orbit multiplet of a hydrogen-like atom, the total magnetic moment is proportional, by Wigner-Eckart theorem, to the total momentum $\mathbf{J}=\mathbf{L}+\mathbf{S}$.

The spin part is critical because $\Delta m_S = \pm 1$ in combination with $\Delta L = \pm 1$, transitions are only excited through the toroidal coupling, and not through electric or magnetic dipole moments.

**DISCUSSION**

We now show that toroidal dipole transitions can be observed in hydrogen-like atoms when the spin projection $m_s$ is a good quantum number. In practice, this means Paschen-Back regime: strong static magnetic field and weak fine structure interaction. We will skip the consideration of nuclear spin because the transitions in question conserve nuclear spin orientation.

Consider a hydrogen-like atom exposed to a plane electromagnetic wave with circular polarization propagating along the Z axis, here chosen as the quantization axis. $\boldsymbol{B} = B\hat{\epsilon}_\pm$ is the optical magnetic field and $\hat{\epsilon}_\pm$ is a unit vector corresponding to the two possible signs of circular polarization. The interaction Hamiltonian in Eq (8) becomes:

$$\hat{H}_{\text{AME}}\left(\hat{o}_\pm\right) = \mp \frac{\alpha}{\sqrt{2}} \frac{\mu_B^2}{ec} \frac{Z}{r^3} \frac{B}{k} \left(\rho e^{\pm i\varphi} \boldsymbol{\sigma}_Z - z\boldsymbol{\sigma}_\pm\right) \quad (10)$$

where $\varphi$ is the azimuthal angle, $\rho$ the radial position operator, $k$ the wavenumber, and $\boldsymbol{\sigma}_\pm = \boldsymbol{\sigma}_X \pm \boldsymbol{\sigma}_Y$. The last term is of particular interest because $z\boldsymbol{\sigma}_\pm$ simultaneously flips the spin and drives spatial orbital transitions with $\Delta L = \pm 1$.

Consider now $n^2S_{1/2} \to n'^2P_{3/2,1/2}$ transition for which the energies are shown in Fig. 2 as functions of the external static magnetic field $\boldsymbol{B}_{DC} = B_{DC}\hat{z}$. The transitions $|m = 0, m_s = \pm 1/2\rangle_g \to |m = 0, m_s = \mp 1/2\rangle_e$ (black dashed arrows) are only toroidal dipole allowed.

In a zero magnetic field, spin-orbit coupling partially lifts the excited state degeneracy and makes $j$ a good quantum number (*23*). This situation is not suitable for observation of toroidal transitions because toroidal transitions then coincide in energy with the much stronger electric dipole ones. However, in strong magnetic fields, orbital and spin quantum numbers again become independent, and toroidal transitions become distinguishable.

State mixing (green arrows in Fig. 2) decreases as $\varepsilon = \Delta E_F / 2\mu_B B_{DC}$, where $\Delta E_F$ is the fine structure energy splitting; toroidal transition rate exceeds electric dipolar transition rate when:

$$\frac{|\langle 0, \mp 1/2 | H_{\text{AME}} | 0, \pm 1/2 \rangle|^2}{\varepsilon^2 |\langle \mp 1, \pm 1/2 | H_E | 0, \pm 1/2 \rangle|^2} \approx \frac{1}{2} \left(\frac{\alpha a_0 k}{\varepsilon}\right)^2 > 1 \quad (11)$$

where $a_0$ is the Bohr radius, and $H_E = -\boldsymbol{d} \cdot \boldsymbol{E}$ is the electric dipole Hamiltonian. Thus, the most easily observable toroidal transitions are likely to be those with small $\Delta E_F$. For hydrogen-like atoms $\Delta E_F \approx \alpha^4 mc^2 Z^2 / 4n'^3$, which is favourable for light atoms and states with high principal quantum numbers. For actual hydrogen at $B_{DC} = 5$ T, the inequality in Eq (11) is fulfilled for the $n' \approx 51$ Rydberg state and above, which may be addressed from $n = 2$ (Balmer series) with an excitation wavelength around 364 nm. This wavelength is available from standard second harmonic generation laser setups; the fundamental mode may be phase-locked on an optical frequency comb to get a precise and tuneable optical frequency reference.

Although its frequency is now different, the toroidal transition moment is around $\frac{1}{2}(\alpha a_0 k)^2 \approx 2 \cdot 10^{-11}$ times smaller than the electric dipole transition moment. We therefore propose the following practical measures to make its observation practical:

1. Frequency-modulation spectroscopy should be used to detect the weak resonance against the background signal coming from the off-resonance nearby electric dipole transitions (*24*)

2. Because, de-excitation occurs mainly over inelastic electric dipole allowed transitions, appropriate frequency filtering of the fluorescence signal should be performed to remove the spurious signal at the excitation frequency.

3. Hydrogen atoms could be replaced by lithium atoms which also have favourable fine structure splitting (*25*). There, we can consider excitations from the fundamental ground state, but deeper into the UV part of the spectrum.

4. Toroidal dipole coupling operator is odd under time reversal symmetry. It therefore changes sign under a change of the static magnetic field sign (equivalent to a flip in the light polarization state in Eq (10)), whereas the electric dipole operator remains unaffected. Hence, the experiment may be conducted in a differential mode, where the toroidal contribution is extracted from the difference between $B_{DC} > 0$ and $B_{DC} < 0$.

A similar symmetry argument was made in the early days of weak interaction research – by Yakov Zeldovich, who understood that a nucleus with a spin must also have a toroidal dipole moment (*26*). For nuclei, this was later confirmed by measuring the anapole moment (a combination of electric and toroidal dipole moments) of caesium nuclei using optical methods (*12*).

**Acknowledgments:** This work was supported by the UK's Engineering and Physical Sciences Research Council (Grant Nos. EP/M009122/1), the Singapore Ministry of Education [Grant No. MOE2016-T3-1-006 (S)] and [Grant No. MOE-T2EP50120-0005].

**Authors contributions:** IK provided the quantum-mechanical description of toroidal spectroscopy; DW evaluated conditions for the applicability of toroidal spectroscopy; NIZ conceived the idea; all co-authors discussed the results and contributed to writing the manuscript.

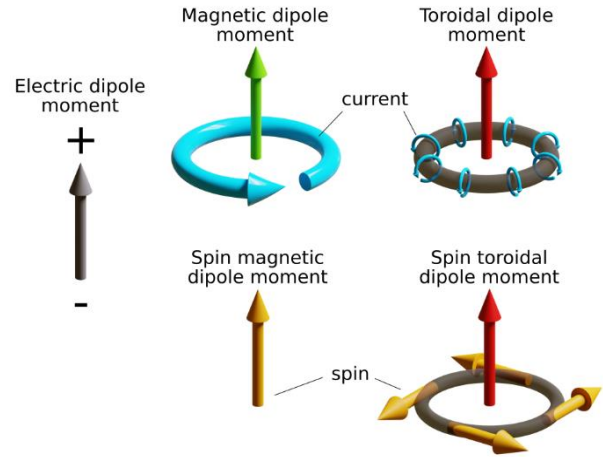

**Fig. 1. Schematic illustrations of static electric, magnetic, and toroidal dipoles in classical electrodynamics**. In relativistic quantum physics, apart from magnetic and toroidal moments induced by charge currents, spin must be considered because it can also contribute to the toroidal dipole moment.

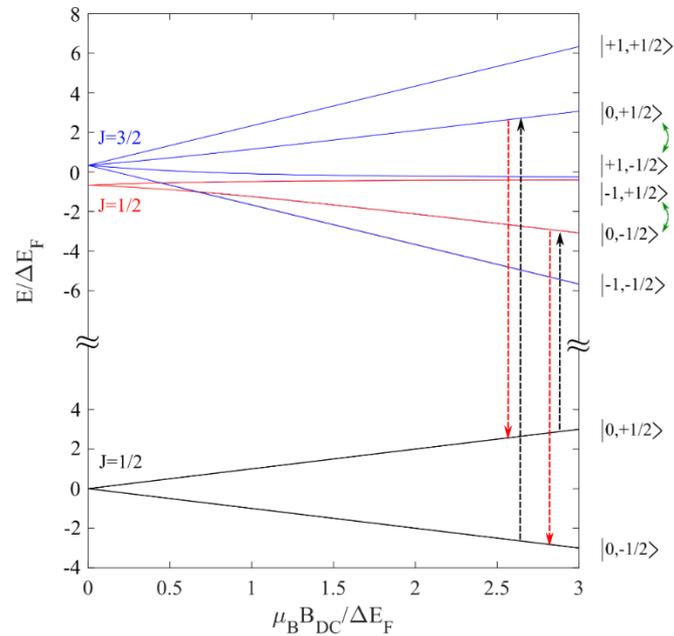

**Fig. 2. Energy spectrum of the transition $n^2S_{1/2} \to n'^2P_{3/2,1/2}$ as a function of the static magnetic field $B_{DC}$.** Black, red and blue curves correspond to the ground state, to $J = 1/2$ excited state manifold, and $J = 3/2$ excited state manifold in the $B_{DC} \to 0$ limit, respectively. Black dashed arrows indicate the toroidal transition; red dashed arrows indicate the electric dipole allowed primary de-excitation route. The quantum numbers in brackets refer to orbital angular momentum (first) and spin (second) projection, in the $B_{DC} \to +\infty$ limit. Green double arrows indicate state mixing due to fine structure coupling.

**Table 1. Selection rules for electric, magnetic, and toroidal dipolar transitions in a hydrogen-like atom**. $L$, $m_L$, and $m_S$ are respectively the orbital quantum number, the orbital projection quantum number, and the spin projection quantum number. For magnetic and toroidal dipole moments, the photon polarization either changes the orbital angular momentum or flips the spin, and therefore $\Delta m_J \leq 1$ (*22*).

| Moment | Parity symmetry | Time reversal symmetry | Orbital selection rule | Spin selection rule | $J$ selection rule |
|---|---|---|---|---|---|
| **Electric dipole** | Odd | Even | $\Delta L = \pm 1$, $\Delta m_L = 0, \pm 1$ | $\Delta m_s = 0$ | $\Delta J = 0, \pm 1$, $\Delta m_J = 0, \pm 1$ |
| **Magnetic dipole** | Even | Odd | $\Delta L = 0$ $\Delta m_L = 0, \pm 1$ | $\Delta m_s = 0, \pm 1$ | $\Delta J = 0, \pm 1$, $\Delta m_J = 0, \pm 1$ |
| **Toroidal dipole** | Odd | Odd | $\Delta L = \pm 1$, $\Delta m_L = 0, \pm 1$ | $\Delta m_s = 0, \pm 1$ | $\Delta J = 0, \pm 1$ $\Delta m_J = 0, \pm 1$ |